\input harvmac
\overfullrule=0pt
\def\Title#1#2{\rightline{#1}\ifx\answ\bigans\nopagenumbers\pageno0\vskip1in
\else\pageno1\vskip.8in\fi \centerline{\titlefont #2}\vskip .5in}

\font\ticp=cmcsc10
%
%
\def\rp{r_+^2}
\def\rmn{r_-^2}
\def\half{{1\over2}}

\def\a{\alpha}

\def\s{{\rm sinh}^2\alpha }
\def\c{{\rm cosh}^2\alpha }

\def\e{\epsilon}

\def\[{\left [}
\def\]{\right ]}
\def\({\left (}
\def\){\right )}
%
%
\lref\dbr{J. Polchinski, S. Chaudhuri, and C. Johnson, hep-th/9602052.}
\lref\jp{J. Polchinski, hep-th/9510017.}
\lref\dm{S. Das and S. Mathur, hep-th/9601152.}
\lref\witb{E. Witten, hep-th/9510135.}
\lref\vgas{C. Vafa, hep-th/9511088.}
\lref\bsv{M. Bershadsky, V. Sadov and C. Vafa,
hep-th/9511222.}
\lref\vins{C. Vafa, hep-th/9512078.}
\lref\cvetd{M. Cvetic and D. Youm, hep-th/9507090.}
\lref\chrs{D. Christodolou, Phys. Rev. Lett. {\bf 25}, (1970) 1596;
D. Christodolou and R. Ruffini, Phys. Rev. {\bf D4}, (1971) 3552.}
\lref\cart{B. Carter, Nature {\bf 238} (1972) 71.}
\lref\penr{R. Penrose and R. Floyd, Nature {\bf 229} (1971) 77.}
\lref\hawka{S. Hawking, Phys. Rev. Lett. {\bf 26}, (1971) 1344.}
\lref\sussc{L.~Susskind,  Phys. Rev. Lett. {\bf 71}, (1993) 2367;
L.~Susskind and L.~Thorlacius, Phys. Rev. {\bf D49} (1994) 966;
L.~Susskind, ibid.  6606.}
\lref\polc{J. Dai, R. Leigh and J. Polchinski, Mod. Phys.
Lett. {\bf A4} (1989) 2073.}
\lref\ascv{A. Strominger and C. Vafa, hep-th/9601029.}
\lref\hrva{P. Horava, Phys. Lett. {\bf B231} (1989) 251.}
\lref\cakl{C. Callan and I. Klebanov, hep-th/9511173.}
\lref\prskll{J. Preskill, P. Schwarz, A. Shapere, S. Trivedi and
F. Wilczek, Mod. Phys. Lett. {\bf A6} (1991) 2353. }
\lref\sbg{S. Giddings, Phys. Rev {\bf D49} (1994) 4078.}
\lref\cghs{C. Callan, S. Giddings, J. Harvey, and A. Strominger,
Phys. Rev. {\bf D45} (1992) R1005.}
\lref\cvyo{M. Cvetic and D. Youm, hep-th/9507090.}
\lref\bhole{G. Horowitz and A. Strominger,
Nucl. Phys. {\bf B360} (1991) 197.}
\lref\bekb{J. Bekenstein, Phys. Rev {\bf D12} (1975) 3077.}
\lref\hawkb{S. Hawking, Phys. Rev {\bf D13} (1976) 191.}
\lref\wilc{P. Kraus and F. Wilczek, hep-th/9411219, Nucl. Phys.
{\bf B433} (1995) 403. }
\lref\ght{G. Gibbons, G. Horowitz, and P. Townsend, hep-th/9410073,
Class. Quantum Grav.,
{\bf 12} (1995) 297. }
\lref\intrp{G. Gibbons and P. Townsend, Phys. Rev. Lett.
{\bf 71} (1993) 3754.}
\lref\gmrn{G. Gibbons, Nucl. Phys. {\bf B207} (1982) 337;
G. Gibbons and K. Maeda Nucl. Phys. {\bf B298} (1988) 741.}
\lref\bch{J. Bardeen, B. Carter and S. Hawking,
Comm. Math. Phys. {\bf 31} (1973) 161.}
\lref\mdr{W. Zurek and K. Thorne, Phys. Rev. Lett. {\bf 54}, (1985) 2171.}
\lref\stas{A.~Strominger and S.~Trivedi,  Phys.~Rev. {\bf D48}
 (1993) 5778.}
\lref\jpas{J.~Polchinski and A.~Strominger,
hep-th/9407008, Phys. Rev. {\bf D50} (1994) 7403.}
\lref\send{A. Sen, hep-th/9510229, hep-th/9511026}
\lref\cvet{M. Cvetic and A. Tseytlin, hep-th/9512031.}
\lref\kall{R. Kallosh, A. Linde, T. Ortin, A. Peet andA. van Proeyen, Phys. Rev. {\bf D46} (1992) 5278.}
\lref\lawi{F. Larsen and F. Wilczek, hep-th/9511064.}
\lref\bek{J. Bekenstein, Lett. Nuov. Cimento {\bf 4} (1972) 737,
Phys. Rev. {\bf D7} (1973) 2333, Phys. Rev. {\bf D9} (1974) 3292.}
\lref\hawk{S. Hawking, Nature {\bf 248} (1974) 30, Comm. Math. Phys.
{\bf 43} (1975) 199.}
\lref\cama{C. Callan and J. Maldacena, hep-th/9602043.}
\lref\sen{A. Sen, hep-th/9504147, Mod. Phys. Lett. {\bf A10} (1995) 2081.}
\lref\suss{L. Susskind, hep-th/9309145.}
\lref\sug{L. Susskind and J. Uglum, hep-th/9401070, Phys. Rev. {\bf D50}
 (1994) 2700.}
\lref\peet{A. Peet, hep-th/9506200.}
\lref\tei{C. Teitelboim, hep-th/9510180.}
\lref\carl{S. Carlip, gr-qc/9509024. }
\lref\thoo{G. 'tHooft, Nucl. Phys. {\bf B335} (1990) 138
Phys. Scr. {\bf T36} (1991) 247.}
\lref\fks{S. Ferrara, R. Kallosh and A. Strominger, hep-th/9508072,
Phys. Rev. {\bf D 52}, (1995) 5412 .}
%
\Title{\vbox{\baselineskip12pt
\hbox{hep-th/9602051}}}
{\vbox{
\centerline{Counting States of  }
\centerline {Near-Extremal Black Holes }}}

\centerline{{\ticp Gary T. Horowitz and Andrew Strominger}}\vskip.1in
\centerline{\sl Department of Physics}
\centerline{\sl University of California}
\centerline{\sl Santa Barbara, CA 93106-9530}
\bigskip
\centerline{\bf Abstract}
A six-dimensional 
black string  is  considered and its Bekenstein-Hawking entropy 
computed. It is shown that  
to leading order  above
extremality, this entropy precisely counts the number of string states
with the given energy and charges.
This identification implies that Hawking decay of the 
near-extremal
black string can be analyzed in string perturbation theory 
and is perturbatively unitary.  
\Date{}
%

\newsec{Introduction}
Classical general relativity and quantum field theory in curved
spacetime together 
provide a beautiful thermodynamic description of black holes.
As Hawking showed \hawk, black holes radiate thermally at a temperature
$T = \kappa/2\pi$, where $\kappa$ is the surface gravity. The laws of
thermodynamics are obeyed if one assigns an entropy to the black hole
equal to one quarter of the horizon area \refs{\bek,\hawk}. 
However, thermodynamics 
is only an
approximation to a more fundamental description in terms of quantum
states. There have been many efforts to describe these states for 
black holes \refs{\bekb \hawkb \mdr \thoo \suss \sug \tei \sen \carl 
\lawi -\cvet}.
This is a difficult task since
a full description requires a quantum theory of gravity. 

Recently, there has been further progress in this direction. This was
made possible using a new description of solitonic states in string theory 
\refs{\polc  \hrva \jp \witb \send \bsv \vins - \vgas}.
For a particular five-dimensional extremal black hole, one can now explicitly
count the number of corresponding BPS-saturated 
states in the theory with given charges
and show that, for large charge, the number
grows like $e^{A/4}$ where $A$ is the horizon area \ascv. 

In this paper
we will extend this result to slightly excited, nonextremal black holes. We
will show that to first order away from extremality, the number of states
can still be counted microscopically and 
continues to be given by the black hole entropy formula. 
The identification of extremal 
black hole excitations with string states enables one to use 
string perturbation theory to study the Hawking decay of 
near-extremal black holes. In particular, as we will briefly discuss 
in the last section, this implies that Hawking emission is 
a unitary process in string perturbation theory. 

The five-dimensional black hole of \ascv\ is a six-dimensional black string
which winds around a compact internal circle.
It was shown in \ascv\ that the  black hole 
states can be simply described in terms of a number of degrees of
freedom living on the circle. The extremal black hole has two types of charges.
One charge determines the number of degrees of freedom, and the other
determines their right-moving momenta. The left-moving momenta is 
zero at extremality. One might expect that nonextremal black holes should 
correspond to keeping the same degrees of freedom, but now giving 
them left-moving momenta as well.
This is exactly what we find. 

For our purposes it is clearer to use the six-dimensional black string 
description rather than the five dimensional black hole.
In section 2 the required black string solution 
is discussed. The extremal
solution with zero momentum has zero horizon area, indicating a nondegenerate
ground state. If one adds right-moving momenta, the black string solution
stays extremal, but the horizon area grows with the momenta.
Dimensional reduction of this black string  along its length, reproduces 
(a slight generalization of) the
five-dimensional extremal  black hole in \ascv. If one adds both left and
right moving momenta, the black string becomes nonextremal, and it reduces
to a nonextremal black hole. In section 3, we show that the
number of string states agrees precisely with that given by the black string
entropy.
We conclude with a brief discussion of the implications of this result
in section 4.

\newsec{A General Black String Solution}

Type IIB string theory in six dimensions contains the terms
\eqn\fds{{1\over 16 \pi}
\int d^6x \sqrt{- g}\( R-(\nabla \phi )^2
-{1 \over 12} e^{2\phi}H^2\)}
in the six-dimensional Einstein frame. $H$ denotes the RR three form field 
strength. 
We adopt conventions in
which $G_N=1$. We wish to consider black string 
solutions to \fds, for which 
the line element can be written in the form
\eqn\lml{ds_6^2=e^{2D}(dx_5+A_\mu dx^\mu)^2 +ds_5^2}
where $\mu, \nu=0,1,...4$. $D$ and $A_\mu$ depend only on $x^\mu$, and 
$D$ tends to zero far from the string. 
Nonzero $A_\mu$ is required when the string carries 
longitudinal momentum.  
It is convenient to periodically identify
$x_5 \sim x_5+L$, so that the string winds along a compact dimension of 
asymptotic 
length $L$, which we take to be very large or infinite. The equations of motion following 
from \fds\ are equivalent to those of the five-dimensional action
\eqn\fdde{{L\over 16 \pi}\int d^5x \sqrt{- g}e^D\(
R-(\nabla \phi )^2-{2 \over 3}(\nabla D)^2-{e^{-2D+2\phi} \over 4}H_+^2
-{e^{-2D-2\phi} \over 4} H_-^2-{e^{2D}
 \over 4}G^2\).}
This action contains three $U(1)$ gauge fields:
$G=dA$ is the usual Kaluza-Klein field strength, $H_+$ derives
from the reduction  $H=H_+\wedge dx^5$ (i.e. $(H_+)_{\mu\nu} = H_{\mu\nu 5}$) 
and $H_-=e^{2\phi+D}*H$ where $*$ denotes the {\it five}-dimensional dual.

The six-dimensional string can can carry electric charge with respect to both
$H_+$ and $H_-$,
\eqn\qhd{\eqalign{Q_+&\equiv {1\over 8}\int_{S^3} e^{-D+2\phi}\ *H_+ ,\cr
 Q_- &\equiv {1\over 4\pi^2}\int_{S^3} e^{-D-2\phi}\ *H_-  .\cr}}
It may also carry total ADM momentum $P$ which appears in five dimensions
as the charge associated with $G$:
\eqn\mmt{P\equiv {2  \pi n \over L}={L\over 16 \pi}\int_{S^3} e^{3D}\  *G .}
We have chosen our conventions so that $n$ and $Q_-Q_+ \equiv \half Q^2$ are 
integers\foot{In the notation of \ascv, $n=Q_H$ and $Q^2=Q_F^2$. The field
normalization used here differs from \ascv.}.
For finite momentum density and large $L$, $n>>1$.

Black string solutions are characterized by $Q_-$, $Q_+$, $n$, as 
well as their 
energy  density and the asymptotic value of 
$\phi$. We are primarily interested in the black string 
entropy which cannot depend on the asymptotic value of the 
$\phi$ \refs{\fks,\cvyo,\lawi}. For a special asymptotic value 
$\phi_h$, the sources for $\phi$ (namely $H_-^2$ and $H_+^2$) 
cancel exactly and the equations of motion imply $\phi$ is  
constant everywhere. This special value is
\eqn\qhds{\eqalign{e^{2\phi_h}&={ 2 Q_+ \over \pi^2 Q_-}
.\cr}}  
In order to compute the entropy it is sufficient to consider the solutions 
with $\phi=\phi_h$. These are obtained by boosting the  
non-extremal, zero-momentum, six dimensional black string solution found
in \ght . The result is 
\eqn \qhs{\eqalign{\phi&=\phi_h,\cr
 e^{2\phi-D} \ *H_+&= {4Q_+ \over \pi^2}\e_3  ,\cr
 e^{- 2\phi-D}\ *H_-&= {2Q_- } \e_3  ,\cr
ds^2&= -\[1-\({\rp\c - \rmn \s \over r^2}\)\]dt^2+ \cr
&~~+{\rm sinh}2\a {\rp- \rmn \over r^2}dtdx_5 +
\[1-\({\rmn\c- \rp \s \over r^2}\)\]dx_5^2 \cr 
&~~+\(1-{\rmn \over r^2} \)
^{-1}\(1-{\rp \over r^2}\)
^{-1}dr^2
+r^2d\Omega^2_3,\cr} } 
where $\e_3$ is the volume form on the unit three-sphere, and 
$\a$ is the boost parameter. The parameters $r_\pm$ denote the event
horizon and the inner horizon, and are related to the charge 
by $Q^2\equiv 2 Q_+ Q_- = (\pi r_+  r_-)^2 $. The fields $D$ and 
$A_\mu$ can be read off
by comparing the metric to \lml. The total ADM momentum can be computed
and expressed in terms of the integer $n$ \mmt\ with the result 
\eqn\monn{n={L^2\over 16  }{\rm sinh}2\a  (\rp-\rmn) .}
The ADM energy of these solutions is
\eqn\eadm{E={L \pi \over 8}\[2(\rp +\rmn)+(\c+\s)(\rp-\rmn)\].}
The Hawking temperature is
\eqn\th{T_H={\sqrt{\rp-\rmn} \over 2 \pi \rp }.}   
The associated entropy is 
\eqn\sasso{S={A \over 4}= \half L \pi^{2}r_+^2 {\rm \cosh}\a \sqrt{\rp-\rmn}.}

Extremal solutions can carry all three charges 
$Q_-$, $Q_+$ and $P$, but 
have $T_H=0$ and a double horizon with $ r_+ =r_-$.
Such solutions are obtained from the general family of solutions \qhs\
by taking the limit $r_+ \to r_-$ with $P$ held fixed, 
which requires $\a \to \infty$.
The resulting solutions have energy
\eqn\eext{E_{ext}={L Q \over 2}+{2 \pi n \over L}.}
and entropy \ascv\
\eqn\sext{S=\pi Q\sqrt{2n}.}
It is important to note that when $n=P=0$ the horizon still has small 
curvature 
of order $1/Q$ \ght. Hence $\alpha '$ corrections will remain negligible.
The area of the horizon vanishes because of longitudinal 
contraction along the string. 

We wish to consider solutions which correspond to low-lying, low-temperature  
excitations of the zero-momentum
black string groundstate. 
These are obtained by keeping $\alpha$ finite and expanding
\eqn\xtr{r_\pm= r_0 \pm \e, ~~~\e <<1~, }
where $r_0 = \sqrt{Q/\pi}$.
The ADM excitation energy $\delta  E$ above 
the groundstate energy ($ L Q/2 $), depends on $\e$ to leading order as 
\eqn\dadm{\delta  E ={L\pi r_0 \e \over 2}(\c+\s).}
The entropy is given by 
\eqn\snta{S= L\pi^2 r_0^2 {\rm cosh}\a\sqrt{ r_0 \e}.}
This can be rewritten as 
\eqn\snt{S=\pi Q\bigl(\sqrt{2 n_L}+\sqrt{2 n_R}\bigr),}
where the left- and right-moving momenta of the string obey 
$n_R-n_L=n$ and are defined
by 
\eqn\lrdf{\eqalign{n_R&\equiv { L \over 4 \pi }(\delta  E+P),
\cr n_L&\equiv { L \over  4\pi }(\delta  E-P). }}
\snt\ is  a good approximation if both the energy density 
$\delta  E/L$ and momentum density $P/L$ are small. 
$n_L$ and $n_R$ can still be large
if $L$ is large.
\snt\ incorporates the leading non-trivial behavior near 
extremality and describes the low temperature 
black string thermodynamics.  
In the next section we will show that this entropy formula 
is in precise accord
with the statistical entropy obtained from string theory.

\newsec{Counting Black String Microstates}

We wish to count states in string theory with the same mass and charges
as the black string in the previous section.
Since the black string carries RR charges $Q_+,\
Q_-$, we will use
the perturbative description of these states in terms of $D$-branes.
We consider type IIB string theory compactified on $T^4$ or $K3$.
(The following argument is independent of which space is used
in the compactification.)
A single extended RR string, or D-onebrane, in six dimensions carries 
the charge $Q_+=1$. The dual charge $Q_-$ is one for a single 
RR-fivebrane which wraps the internal four dimensions\foot{If the 
internal space is K3 there is an anomalous shift in the $Q_-$ 
charge \bsv\ which can be ignored for large $Q$.}. The extremal 
black string solution of the previous section corresponds to 
a bound state of $Q_+$ RR strings and $Q_-$ RR fivebranes.
Since the four dimensional compact space is assumed small, this bound
state is  a string in six dimensions.
The entropy of such configurations may be 
counted as follows \ascv\ for $Q_-=1$.\foot{The entropy should depend
only on the product $Q_+ Q_-$. Other values of $Q_-, \ Q_+$ can be obtained
by T-duality, but the counting problem is different.}
The strings and fivebranes do not separate in the noncompact 
six-dimensional spacetime, but the $Q_+$ RR strings are free to wander around in
the internal four-dimensional space. This yields 
$4Q_+$ massless bosons, together with their superpartners, 
in the $1+1$ 
effective field theory on the string.\foot{These correspond to 
fundamental open strings 
whose Dirichlet boundary conditions confine them to the string.} 
Extremal 
BPS configurations with nonzero momentum can be obtained by 
exciting only the right-moving components of these 
massless fields. For $n_R >>1$, the number of such states is given by 
the standard two-dimensional entropy-energy relation $S = 2\pi\sqrt{ cn_R/6}$. 
For $2Q^2$ species of fermions and bosons, 
$c= 3Q^2$ and thus
\eqn\fbr{S=\pi Q\sqrt{2n_R}}
in perfect agreement with \snt\ \ascv. This result is valid in 
the thermodynamic limit of large $n$, which can always be attained for any 
fixed momentum density by taking $L$ to be large. 

The non-extremal case is a simple extension of this. Now we must 
drop the restriction to pure right-movers and count states with 
a given $(n_L, n_R)$. At low energies and 
densities the interactions between left and right movers can be ignored
and the statistical entropy is just the sum 
\eqn\sntt{S=\pi Q\bigl(\sqrt{2n_L}+\sqrt{2n_R}\bigr),}  
again in perfect agreement with \snt.  Since the energy of the 
black string is proportional to 
$L$, we can get arbitrarily near extremality 
and remain in the thermodynamic limit by taking $L$ sufficiently large, 
hence avoiding the limitations on the statistical description of near-extremal 
black holes discussed in \prskll. 

\newsec{Discussion}

Our results bear on the issue of unitary in black hole evaporation. 
One can view the process of scattering by an extremal black hole in terms of 
an absorption of the incident quanta (which excites the black hole 
just above extremality) followed by 
Hawking decay back to extremality. Since Hawking radiation  
is involved, it has been argued that information about the 
incident quanta is lost in the black hole, and unitarity is violated. 
However we may alternatively describe this  process in terms of 
string scattering 
by D-branes. This has been understood in some detail recently
\refs{\polc,\jp,\cakl} and is certainly unitarity. Hence 
perturbative string theory provides a unitary description of scattering 
off certain extremal black holes.

However this does not resolve the issue of information loss
for the following reason \refs{\ascv, \dbr}.  The ratio of the string 
length to the Planck length grows as an inverse power of the string coupling. 
The size of an extremal RR black hole (as measured by its Schwarzschild radius) 
is a power of its charge times
the Planck length. Hence in string perturbation theory strings are treated as 
much larger than the RR black holes. The perturbative stringy description 
of a RR 
black hole is as a D-brane with no analog of an
event horizon.
Proponents of 
unitarity violation 
might argue that it is not surprising that 
a description which can not see the event horizon also can not see 
information loss.  As an analogy,  perturbative unitarity of 
flat-space graviton scattering in string theory
seems to be universally accepted, 
yet Hawking has argued that non-perturbatively black holes can be formed 
and unitarity will be violated. 
Further exploration of these issues is certainly in order.

The D-brane description is generally valid only for 
very weak coupling, $g_s< 1/Q^2$, because open string loops 
couple proportionally to the number of D-branes. 
At stronger coupling the Schwarzschild radius becomes larger than
the string scale. In this regime, the D-brane description is unreliable 
and the black hole description is valid. Given that the two 
descriptions do not appear to have an overlapping region of validity,
one may wonder 
why our two calculations, which utilize different descriptions, 
are in agreement. In the extremal case discussed in \ascv, 
the topological stability of BPS states was used to 
argue that the results of the D-brane calculation could be 
extrapolated 
from weak to strong coupling. That argument is not directly 
applicable here because the 
states under consideration are not all BPS-saturated. However 
similar reasoning can be applied. The entropy computed in the 
D-brane picture is independent of the coupling to first order above extremality. 
That is because,
for very large $L$, we 
are considering only long wavelength modes with
very small energy densities and correspondingly small interactions. 
As the coupling is 
turned up, interactions between left- and right-moving modes of the 
D-brane become stronger. Nevertheless, since the leading-order 
low-energy result \snt\
is coupling-independent, one expects the answer to change only if there is 
a phase transition which changes the number of degrees of freedom. 
There is no reason to expect such a transition to occur, and 
the remarkable agreement between the weak-coupling D-brane 
result and the strong-coupling black hole result is evidence 
that it does not occur. Going beyond the calculation described here - 
for example to numerically compare the decay rate or S-matrices 
computed in the black hole and D-brane pictures - may well 
require grappling with the problem of strong coupling. 
Perhaps string duality will be useful in this regard.

\centerline{\bf Acknowledgements}
We would like to thank M. Cvetic, D. Lowe, J. Polchinski, C. Vafa 
and E. Witten for useful discussions. 
Some observations related to those of this paper have recently been 
made independently in \refs{\dm, \cama}.
The research of G.H. is supported in part by NSF Grant PHY95-07065
and that of A.S. is supported in part by DOE grant DOE-91ER40618.

\listrefs
\end